\documentclass[letterpaper]{article}
\usepackage{latexsym,amsfonts,amsmath,amsthm,amssymb}
\usepackage{pstricks}

\begin{document}

\title{Quantum-like Representation of Macroscopic Configurations}
\author{Andrei Khrennikov\\
International Center for
Mathematical Modeling \\ in Physics and Cognitive Sciences\\
University of V\"axj\"o, S-35195, Sweden}

\maketitle

\abstract{The aim of this paper is to apply a contextual probabilistic model (in the spirit of Mackey, 
Gudder, Ballentine) to represent and to generalize some results of quantum logic 
about possible macroscopic quantum-like (QL) behaviour. The crucial point is that our model provides 
QL-representation of macroscopic configurations 
in terms of complex probability amplitudes -- wave functions of such configurations. Thus, instead of the language 
of propositions which is common in quatum logic, we use the language of wave functions which 
is common in the conventional presentation of QM.  We propose
a quantum-like representation algorithm, QLRA, which maps probabilistic data of any origin in
complex (or even hyperbolic) Hilbert space. On the one hand, this paper clarifyes some questions in foundations of
QM,  since some
rather mystical quantum features are illustrated on the basis of behavior of macroscopic systems. 
On the other hand, the approach developed in this paper may be used e.g. in biology, sociology, or psychology.
Our example of QL-representation of hidden macroscopic configurations can find natural applications in those
domains of science.}


\section{Introduction}
One should sharply distinguish QM as a physical theory and the mathematical formalism of QM. In the same way 
as one should distinguish classical Newtonian mechanics and its mathematical formalism. Nobody is surprised
that the differential and integral calculi which are basic in Newtonian mechanics can be fruitfully applied in other 
domains of science. Unfortunately, the situation  with the mathematical formalism of QM is essentially 
more complicated --
some purely mathematical features of QM are  identified with features 
of quantum physical systems. Although already Nils Bohr pointed out \cite{Bohr}, see also \cite{PL1},
\cite{PL2},  to the possibility 
to apply the mathematical formalism of QM outside of physics, prejudice based on the identification of 
mathematics and physics still survives (but cf. e.g. Accardi, Aerts,  Ballentine, De Muynck, 
Grib et al., 
Gudder, Gustafson,  Land\'e, Mackey 
\cite{AC}--\cite{MC1} and also \cite{FPP}--\cite{FPP3}). 
One can point out just to a few applications outside of physics. Here we discuss not 
{\it reductionist models} in that the quantum description appears as a consequence 
of the evident fact that any physical system, even living (for example, the brain, see e.g. \cite{P1}, \cite{P2}), 
is composed of quantum particles, 
but really the possibility to use the mathematical 
formalism of QM without direct coupling with quantum physics, see e.g. \cite{AR}, 
\cite{KH1}, \cite{KH2}, \cite{GR1}--\cite{GR3}.     

We remark that importance of mentioned separation between quantum physics and quantum mathematics has been already well
recognized in {\it quantum logic,} see e.g. Mackey \cite{MC1} or Beltrametti and Cassinelli \cite{BELT}. 
In particular,
an exiting possibility to apply quantum mathematics to macroscopic systems is not surprising for quantum logicians.
 However, one could not see
visible results of  diffussion of this quantum logic knowledge into real quantum physics. There are a few reasons
for this, in particular, psychological ones. It seems that the main problem is that the majority of 
physicists think that QM is not about  new logic, but new physics. 
Thus the massage of Birkhof and von Neumann\cite{BV}, as well as 
Bohr \cite{Bohr} who 
discussed a possibility to reduce quantum particularities to elaboration of new ``quantum language'', was practically 
ignored in quantum physics. 

We point out that quantum logic is closely interrelated with {\it quantum probability  which is
a caculus of complex probability amplitudes and self-adjoint operators} (in contrast to classical
Kolmogorovian probability theory which is a calculus of measures and measurable functions, random variables). 
Roughly speaking quantum logic emphasizes the observational part quantum formalism, the calculus of 
propositions  \cite{BELT}
representing results of quantum observations. The complex probability amplitude (the wave function)  
does not belong to 
the main 
field of interest of quantum logicians.\footnote{Personally 
I am not quantum logician. Thus my interpretation
may be wrong from the internal viewpoint of quantum logic. 
But I think that it has right as rather general
external opinion.} On the other hand, the wave function is the basic object of practical quantum 
physics \cite{D}, \cite{LANDAU}.

Recently I developed so called contextual probability theory \cite{KHC} which was inspired essentially by quantum logic
and quantum probability, especially Mackey's approach \cite{MC1}.\footnote{I was lucky to meet George Mackey 
at the Congress of Quantum Structures Association
in  Castiglioncello, Italy, 1992. Our conversations on the probabilistic structure of QM 
were the starting point of my further studies on contextual
probability. I  also was lucky to speak shortly  with Andrei Nikolaevich Kolmogorov
(when he submited my paper to Doklady Akademii Nauk USSR). I was surprised that personally 
he was not satisfied by his axiomatics of probability theory \cite{KOL}. Later (after his death) 
his former 
students Albert Shiryaev and Alexander Bulinskii explained me that contextuality of probabilities
and the impossiblity to play the whole game with a single Kolmogorov space 
was evident for Kolmogorov \cite{KOL1}.}
The main distinguishing feature of the theory of contextual probabilities is a 
possibility to derive the complex probability amplitude, 
the wave function, from probabilistic data.  Such 
an algorithm for mapping of probabilistic data into the 
complex probability amplitude was proposed in \cite{KHC}, 
{\it quantum-like representation algorithm} -- QLRA.
This algorithm also generates representation of observables (in fact, to 
fixed ``reference observables'') by self-adjoint operators.
Thus by contextual probability theory the mathematical quantum structure is not fundamental. 
It appears as a special representation 
of probabilistic data. The main distinguishing feature of the QL-representation
is {\it ignorance by details about system's behaviour} which are not approachable
by an external observer. This is a {\it consistent} way to proceed within
 incomplete description of system's behavior.\footnote{Consistency is an 
extremely important feature of quantum and QL representations of probabilistic data.
Of course, one may try cut off data ocationaly, but such a data-processing would (soon or later)
induce chaos.}

The aim of this paper is to use our contextual probabilistic model, the {\it V\"axj\"o model,}
to represent and to generalize some results of quantum logic on macroscopic quantum-like (QL) behaviour
in terms of complex probability amplitudes. On the one hand, it may be intersting for physicists, since some
rather mystical quantum features will be illustrated on the basis of behavior of macroscopic systems. 
On the other hand, the approach developed in this paper may be used e.g. in biology, sociology, or psychology.
Our example of QL-representation of hidden macroscopic configurations can find natural applications in 
these domains of science.	

The basic example which we would like to generalize in 
the  contextual probabilistic framework is well known in quantum logic.
This is ``firefly in the box''. It was proposed by Foulis who wanted to show that a macroscopic system, firefly,
can exhbit a QL-behavior which can be naturally represented in terms of quantum logics. First time this example
was published in Cohen's book \cite{Cohen}, a detalied presentation can be found in Foulis' paper \cite{Foulis}, 
see also Svozil \cite{Svozil}. Later ``firefly in the box''
was generalized to a so called  generalized urn's model, by Wright \cite{ Wright} (psychologist).      

From the viewpoint of quantum logic such examples illustrate the following problem. For a given quantum logic
one wants to find a Boolean algebra such that by ignoring some elements of this algebra one obtains 
the original quantum  logic. I would formulate this problem in the following way: ``To quantum (and more general
QL) structures through ignorance of some information about underlying classical Boolean algebras.''
 We shall use
two lessons of previous studies in quantum logic: a) essentially quantum structures (lattices of quantum projectors) can be obtained 
from purely classical Boolean models; b) not all quantum structures have underlying classical Boolean models.

Similar lessons we have from studies on contextual probability: 

\medskip

a). The QLRA can be applied to classical probabilistic data (which can be described by the Kolmogorov 
model). The result will be nontrivial: Born's rule, interference of probabilities, representation of Kolmogorovian
random variables by self-adjoint operators. Thus all basic 
quantum structures are present in the classical probabilistic models, but in  a latent form.

\medskip

b). The quantum probabilistic structure could not be completely reproduced on the basis of a single Kolmogorov
probability space (by using Gudder's terminology one must consider a {\it probability manifold} with the 
atlas consisting of  a few Kolmogorovian charts).     

\medskip

Regarding b) we point out to one very important difference between quantum logic and contextual probability theory.
According to the latter even in the two dimensional case an undelying classical model does not exist. By applying QLRA
to probabilistic data obtained on the basis of a single Kolmogorov probability space we are not able to get 
{\it all pure quantum states and all pairs of noncommutative observables.} To show this, we use an analogue 
of Bell's inequality for transition probabilities, see \cite{CBI} and appendix.

In general, our contextual model is based on the {\it frequency definition of probability} which was formalized
by R. von Mises \cite{MI} (this formalization was simplified and justified in \cite{KH3}). By using frequency probabilities 
we can reproduce completely the pobabilistic structure of QM. However, the contextual statistical model is not reduced
to the quantum probabilistic model. Besides ordinary trigonometric $\cos$-interference it predicts 
hypebolic $\cosh$-interference. Corresponding contexts are repsented not by complex, but hyperbolic probability 
amplitude, i.e., in an analogue of Hilbert space, but over the algebra of hypebolic numbers,
$z=x+jy, j^2=+1, x, y \in {\bf R},$ see  \cite{KHC}.

We can mention some consequences of our  QL-representation of macroscoipic configurations 
for foundations of quantum physics. 
All distinguishing features of the quantum probabilistic behavior can be modeled by using
macroscopic systems. For such macroscopic models the QL-description is not complete.
Thus hidden variables exist, but they could not be observed on the basis of available observables.
Those observables which we (external observers) could use are too fuzzy, cf. \cite{Lahti}. 
Nevertheless,  a kind of
 {\it  Einstein's demon}  can observe behavior of hidden variables.\footnote{This demon is similar
to Maxwell's demon who might (in principle) violate the principles of thermodynamics. Einstein's 
demon might violate principles of the Copenhagen interpretation of QM, in particular, the principle
of complementarity.}
 Since our examples are macroscopic,
 such  Einstein's demon  can be a macroscopic observer.
Classical probability describes models in that measurements of complementary observables are 
not mutually disturbing. As we remarked, such models do not cover completely QM.\footnote{To show this, we use
an analogue of Bell's inequality which we obtained for transition probabilities, see appendix.} 
On the other hand, by using 
models with mutual distrbance and the frequency approach to probability we can reconstruct QM in the
realistic framework. We also discuss ``fly-realization'' of the EPR-Bohm experiment. Since flyes are 
macroscopic systems, realism could not be questioned. Possible explanations of violation of Bell's inequality
are nonlocality \cite{Bell}, unfair sampling \cite{Santos}, \cite{AD}, \cite{AD1}, 
ensemble nonreproducibility \cite{DEB}, 
\cite{KH3}, \cite{KHRT}, \cite{HPL2}. 
For macroscopic systems
the latter two possibilities are essentially
more natural than the first one.

Of course, we understand well that our {\it fly-methaphor} can not be used for 
derivation of crucial consequences about 
microscopic quantum systems, such as photons and electrons. It might be that similarities in mathematical description
are just occational. Nevertheless, these similarities are really astonishing.

\section{Firefly in the box}

We recall the well known example  \cite{Foulis} of QL-behavior.
We modify its presentation by emphasizing its probabilistic structure.
 Let us consider a box which is divided into 
four sub-boxes. These small boxes which are denoted by 
$\omega_1, \omega_2, \omega_3, \omega_4$ provides
internal description. These elements are avalaible for  Einstein's demon,
but they are not avalaible for some external observable. 

 We consider the Kolmogorov probability space:
$\Omega=\{ \omega_1, \omega_2, \omega_3, \omega_4\},$ the algebra of all finite subsets 
${\cal F}$ of $\Omega$ and a probability measure determined by probabilities 
${\bf P}(\omega_j)=p_j,$ where $ 0< p_j< 1, p_1+...+p_4 = 1.$  

\psset{unit=1cm}

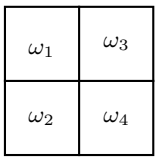
\begin{figure}
\centering
\begin{pspicture}(0,0)(2,2)
\psframe(0,0)(2,2)
\psline(0,1)(2,1)
\psline(1,0)(1,2)
\rput(0.5, 1.3){\parbox{1.75cm} {\raggedright\small
\begin{center}$\omega_1$\end{center}}}
\rput(1.5, 1.5){\parbox{1.75cm} {\raggedright\small \begin{center}$\omega_3$
\end{center}}}
\rput(0.5, 0.5){\parbox{1.75cm} {\raggedright\small \begin{center}$\omega_2$
\end{center}}}
\rput(1.5, 0.5){\parbox{1.75cm} {\raggedright\small \begin{center}$\omega_4$
\end{center}}}
\end{pspicture}
\caption{Internal description.}
\end{figure}

We now consider two different disjoint partitions of the set $\Omega:$ 
$$
C_{\alpha_1}=\{ \omega_1, \omega_2\}, C_{\alpha_2}=\{ \omega_3, \omega_4\},
$$
$$
C_{\beta_1}=\{\omega_1, \omega_4\}, C_{\beta_2}=\{\omega_2, \omega_3\}.
$$
We can obtain such partitions by dividing the box: 
a) into two equal parts by the vertical line:
the left-hand part gives $C_{\alpha_1}$ and the right-hand part $C_{\alpha_2};$
b) into two equal parts by the horizontal line:
the top part gives $C_{\beta_1}$ and the bottom part $C_{\beta_2}.$

We introduce two  random variables corresponding to these partitions:
$\xi_a(\omega) =\alpha_i,$ if $\omega\in C_{\alpha_i}$ and $\xi_b(\omega)=\beta_i\in$ if $\omega\in C_{\beta_i}.$
Here $\alpha_i$ and $\beta_i$ are arbitrary labels. 
Suppose now that the external observer is able to measure only these two variables, denote the corresponding
observables by the symbols $a$ and $b.$ We remark that there exist other 
random variables, they are avalaible
for  Einstein's demon, but not for the external observer.\footnote{For example, $\xi(\omega)=+1,\;
\omega=\omega_1, \omega_2, \omega_3,$ and  $\xi(\omega)=-1,\;
\omega=\omega_4.$}
Roughly speaking elements $\omega_j$ are not visible for the latter observer. They are
``hidden variables.''  

\psset{unit=1cm}

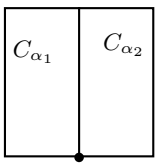
\begin{figure}
\centering
\begin{pspicture}(0,0)(2,2)
\psframe(0,0)(2,2)
\psline(1,0)(1,2)
\rput(0.4, 1.3){\parbox{1.75cm} {\raggedright\small
\begin{center}$C_{\alpha_1}$\end{center}}}
\rput(1.6, 1.5){\parbox{1.75cm} {\raggedright\small \begin{center}$C_{\alpha_2}$
\end{center}}}
\psdot(1, 0)
\end{pspicture}
\caption{The $a$-observable.}
\end{figure}

\begin{figure}
\centering
\begin{pspicture}(0,0)(2,2)
\psframe(0,0)(2,2)
\psline(0,1)(2,1)
\psdot(2,1)
\rput(1, 1.3){\parbox{1.75cm} {\raggedright\small
\begin{center}$C_{\beta_1}$\end{center}}}
\rput(1, 0.5){\parbox{1.75cm} {\raggedright\small \begin{center}$C_{\beta_2}$
\end{center}}}
\end{pspicture}
\caption{The $b$-observable}
\end{figure}
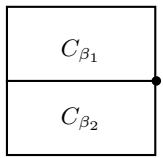

Such a probabilistic model can be illustrated by the following example \cite{Foulis}.  Let us consider
a firefly in the box. It has definite position in space. The firefly position can be seen by 
 Einstein's demon  living inside this box. 

Now we consider an external observer who  has only two possibilities 
to observe the firefly in the box: 

\medskip

1) to open a small window at the point $a$ which is located in 
such a way (the bold 
dot in the middle of the bottom side of the box, Figure 2) that it is 
possible to determine  only either the firefly is in the section $C_{\alpha_1}$ 
or in the section $C_{\alpha_2}$ of the box; 

2) to open a small window at the point $b$ which is located in 
such a way (the bold dot in the middle of the right-hand 
side of the box, Figure 3) that it is possible to determine  only either the firefly is in the section $C_{\beta_1}$ 
or in the section $C_{\beta_2}$ of the box. 

\medskip

In the first case such an external observer can determine in which part, $C_{\alpha_1}$ or $C_{\alpha_2},$
the firefly is located. In the second case he can only determine in which part, 
$C_{\beta_1}$ or $C_{\beta_2},$ the firefly is located. But he is not able to look into both windows
simultaneously. In such a situation the observables $a$ and $b$ 
are the only source of information about the firefly (``reference observables''). 
The Kolmogorov description is meaningless for the external observer (although it is present 
in the latent form), but it is very useful for Eistein's demon. 

Can one apply in such a situation the  QL-description? Can we construct the wave function of 
the firefly in the box? Can we represent observables (in fact, classical random variables)
$a$ and $b$ by self-adjoint operators?  
The answers are to be positive.

\section{Contextual probability}
 
A general statistical model for observables based on the contextual
viewpoint to probability will be presented. It will be shown that
classical as well as quantum probabilistic models can be obtained as
particular cases of our general contextual model, the {\it{V\"axj\"o
model}}, \cite{KHC}. As was mentioned in introduction, I was inspired 
by Mackey's  program: {\it To deduce the probabilistic formalism of 
quantum mechanics starting with a system of natural probabilistic
axioms.} We reduced essentially the number of axioms (Mackey had 8 axioms and we have only two 
axioms). But the main differernce between Mackey's model and the V\"axj\"o model is that Makey
should postulate the complex Hilbert space structure, but in our model it is derived from our 
two axioms. Moreover, representations of the V\"axj\"o
model  are not reduced to the conventional, classical and quantum
ones. Our model also implies
hyperbolic $cosh$-interference that induces  ``hyperbolic quantum
mechanics'' \cite{KHC}.

A physical, biological, social,  mental, genetic, economic, or financial
{\it context}  $C$ is  a complex of corresponding conditions.
Contexts are fundamental elements of any contextual probabilistic  model. Thus construction  
of any model
$M$ should be started with fixing the collection of  contexts of this model.
Denote the collection of contexts
by the symbol ${\cal C}$ (so the family of contexts  ${\cal C}$ is determined by the 
model $M$ under consideration). In the mathematical formalism ${\cal C}$ is an abstract set
(of ``labels'' of contexts).  

We remark that in some models it is possible to construct a set-theoretic 
representation of contexts -- as some family of subsets of a set $\Omega.$ For example,
$\Omega$ can be the set of all possible parameters (e.g., physical, or mental, or economic)
of the model. However, in general we {\it do not assume the possibility to construct a set-theoretic 
representation of contexts.}

Another fundamental element of any contextual probabilistic  model 
$M$ is a set of observables ${\cal O}:$
each observable $a\in {\cal O}$ can be measured
under each complex of conditions $C\in {\cal C}.$  
For an observable $a \in {\cal O},$ we denote the set
of its possible values (``spectrum'') by the symbol
$X_a.$

We do not assume that all these observables can  be measured
simultaneously. To simplify considerations, we shall consider only
discrete observables and, moreover, all concrete investigations will
be performed for {\it dichotomous observables.}

\medskip

{\bf Axiom 1:} {\it For  any observable
$a \in {\cal O}$  and its value $\alpha \in X_a,$ there are defined contexts, say $C_\alpha,$
corresponding to $\alpha$-selections: if we perform a measurement of the observable $a$ under
the complex of physical conditions $C_\alpha,$ then we obtain the value $a=\alpha$ with
probability 1. We assume  that the set of contexts ${\cal C}$ contains 
$C_\alpha$-selection contexts for all observables $a\in {\cal O}$ and $\alpha \in X_a.$}

\medskip

For example, let $a$ be the observable corresponding to some question:  $a=+$ (the answer ``yes'')
and $a=-$ (the answer ``no''). Then the $C_{+}$-selection context is the selection of 
those participants of the experiment who answering ``yes'' 
to this question; in the same way we define the  
$C_{-}$-selection context. By Axiom 1 these contexts are well defined. 
We point out that in principle a participant of this experiment might not want to reply at all 
to this question or she  might change her mind immediately after her answer.
By Axiom 1 such possibilities are excluded. By the same 
axiom both $C_{+}$ and $C_{-}$-contexts belong to the system of contexts under consideration.

\medskip

{\bf Axiom 2:} {\it There are defined contextual (conditional) 
probabilities $p_C^a(\alpha) \equiv {\bf P}(a= \alpha\vert C)$ for any
context $C \in {\cal C}$ and any observable $a \in {\it O}.$}

\medskip

Thus, for any context $C \in {\cal C}$ and any observable $a \in {\it O},$ 
there is defined the probability to observe the fixed value $a=\alpha$ under the complex 
of conditions $C.$

Especially important role will be played by ``transition probabilities''
$
p^{a\vert b}(\alpha\vert \beta)\equiv {\bf P}(a=\alpha\vert C_\beta), a, b \in {\cal O}, \alpha \in X_a, \beta \in X_b,
$
where $C_\beta$ is the $[b=\beta]$-selection context. By axiom 2 for any context $C\in {\cal C},$ 
there is defined the set of probabilities:
$
 \{ p_C^a : a \in {\cal O}\}.
$
We complete this probabilistic data for the context $C$  by transition probabilities.
  The corresponding collection of data $D({\cal O}, C)$ 
consists of contextual probabilities:
$
p^{a\vert b}(\alpha\vert \beta),p_C^b(\beta),
p^{b\vert a}(\beta \vert \alpha), p_C^a(\alpha)...,
$
where $a,b,... \in {\cal O}.$ Finally, we denote  the family of
probabilistic data $D({\cal O}, C)$ for all contexts  $C\in {\cal
C}$ by the symbol ${\cal D}({\cal O}, {\cal C}) 
(\equiv \cup_{C\in {\cal C}} D({\cal O}, C)).$

\medskip

{\bf Definition 1.} (V\"axj\"o Model) {\it A contextual  probabilistic  model of reality is a triple
$M =({\cal C}, {\cal O}, {\cal D}({\cal O}, {\cal C})),$
where ${\cal C}$ is a set of contexts and ${\cal O}$ is a  set of observables
which satisfy to axioms 1,2, and ${\cal D}({\cal O}, {\cal C})$ is probabilistic data
about contexts ${\cal C}$ obtained with the aid of observables belonging ${\cal O}.$}

\medskip

We call observables belonging the set ${\cal O}\equiv {\cal O}(M)$ {\it reference of observables.}
Inside of a model $M$  observables  belonging  to the set ${\cal O}$ give the only possible references
about a context $C\in {\cal C}.$ In the definition of the V\"axj\"o Model we speak about ``reality.''
In our approach it is reality of contexts. 

In what follows we shall consider V\"axj\"o models with two dichotomous reference observables.

\section{Frequency definition of probabilities}

The definition of probability has not yet been specified. In this paper we shall use the frequency definition of probability 
as the limit of frequencies in a long series of trials, von Mises' approach, \cite{MI},  \cite{KH3}. We are aware that this approach 
was criticized a lot in mathematical literature. However, the main critique was directed against von Mises' 
definition of randomness. If one is not interested in randomness, but only in frequencies of trials, then the 
frequency approach is well established, see  \cite{KH3}. 

We consider a set of reference
observables ${\cal O}= \{ a, b \}$ consisting of two observables $a$ and $b.$
We denotes the sets of values (``spectra'') of the reference observables by symbols $X_a$ and $X_b,$
respectively.

Let $C$ be some context. In a series of observations of $b$ (which can be infinite in a mathematical model)
we obtain a sequence of values of $b:$
$x\equiv x(b \vert C) = (x_1, x_2,..., x_N,...), \;\; x_j\in X_b.$
In a series of observations of $a$ we obtain a sequence of values of $a:$
$y\equiv y(a \vert C) = (y_1, y_2,..., y_N,...), \;\; y_j\in X_a.$
We suppose that the {\it principle of the statistical stabilization} for relative frequencies \cite{MI}, \cite{KH3}
holds. This means that the frequency probabilities
are well defined:
$p_C^b(\beta) = \lim_{N\to \infty} \nu_N(\beta; x), \;\; \beta \in X_b;$
$p_C^a(\alpha)= \lim_{N\to \infty} \nu_N(\alpha; y), \;\; \alpha\in X_a.$
Here $\nu_N(\beta; x)$ and $ \nu_N(\alpha; y)$ are frequencies of observations of values
$b=\beta$ and $a=\alpha,$ respectively (under the complex of conditions $C).$

{\bf Remark.} (On the notions of collective and $S$-sequence) R. von Mises considered
in his theory two principles: a) the principle of the statistical stabilization for relative frequencies;
 b) the principle of randomness. A sequence of observations for which
both principle hold was called a {\it collective,} \cite{MI}. 
However, it seems that the validity of the principle
of statistical stabilization  is often enough  for applications. Here we shall use just the
convergence of frequencies to probabilities. An analog of von Mises' theory for sequences of observations
 which satisfy the principle of statistical stabilization
was developed in \cite{KH3}; we call such sequences $S$-{\it sequences.}

Everywhere in this paper it will be assumed that {\it sequences of observations are $S$-sequences},
cf. \cite{KH3} (so we are not interested in the validity of the principle of randomness for sequences
of observations, but only in existence of the limits of relative frequencies).

Let $C_{\alpha},  \alpha\in X_a,$  be contexts  corresponding
to  $\alpha$-filtrations, see Axiom 1.
By observation of $b$ under the context $C_\alpha$ we obtain a sequence:
$x^{\alpha} \equiv x(b \vert C_\alpha) = (x_1, x_2,..., x_{N},...), \;\; x_j \in X_b.$
It is also assumed that for  sequences of observations  $x^{\alpha}, \alpha\in X_a,$
the principle of statistical stabilization for relative frequencies
holds true and the frequency probabilities are well defined:
$p^{b \vert a}(\beta \vert \alpha) = \lim_{N \to \infty} \nu_{N}(\beta; x^{\alpha}), \;\;
\beta \in X_b.$
Here $\nu_N(\beta; x^\alpha), \alpha\in X_a,$  are frequencies of observations of value
$b=\beta$ under the complex of conditions $C_\alpha.$
We can repeat all previous considerations by changing $b\vert a$-conditioning to  $a \vert b$-conditioning.
There can be defined probabilities $p^{a \vert b}(\alpha \vert \beta).$

\section{Quantum-like representation algorithm -- QLRA}

In \cite{KHC} we derived the following formula for interference of probabilities:
\begin{equation}
\label{TFR} p_C^b(\beta) = \sum_\alpha p_C^a(\alpha) p^{b\vert
a}(\beta\vert \alpha) + 2 \lambda(\beta\vert  \alpha, C)
\sqrt{\prod_\alpha p_C^a(\alpha) p^{b\vert a}(\beta\vert \alpha)},
\end{equation}
where the coefficient of interference 
\begin{equation}
\label{KOL6}
\lambda(\beta\vert  a, C) = \frac{p_C^b(\beta) - \sum_\alpha p_C^a(\alpha) p^{b\vert
a}(\beta\vert \alpha)}{2 \sqrt{\prod_\alpha p_C^a(\alpha) p^{b\vert a}(\beta\vert \alpha)}} .
\end{equation}
A similar representation we have for the $a$-probabilities.
Such interference formulas are valid for any collection of contextual 
probabilistic data satisfying the conditions:

\medskip

R1). Observables $a$ and $b$ are symmetrically conditioned\footnote{This condition will induce symmetry 
of the scalar product}:
$$
p^{b\vert a}( \beta \vert  \alpha)= p^{a \vert b}( \alpha \vert  \beta).
$$ 

\medskip

R2). Observables $a$ and $b$ are mutually nondegenerate\footnote{This condition will induce
noncommutativity of operators $\hat{a}$ and $\hat{b}$ representing these observables.}: 
$$
p^{a\vert b}( \alpha \vert  \beta)>0, \; \; p^{b \vert a}( \beta\vert  \alpha)>0.
$$ 

R2a). Context $C$ is nondegenerate with respect to both observables $a$ and $b:$ 
$$
p_C^b(\beta) >0, \; p_C^a(\alpha) >0.
$$
\medskip

Suppose that also the following conditions hold:

\medskip

R3). Coefficients of interference are bounded by one\footnote{This condition will induce representation of the context 
$C$ in the complex Hilbert space. Thus complex numbers appear due to this condition.}:
$$
\Big\vert \frac{p_C^b(\beta) - \sum_\alpha p_C^a(\alpha) p^{b\vert
a}(\beta\vert \alpha)}{2 \sqrt{\prod_\alpha p_C^a(\alpha) p^{b\vert a}(\beta\vert \alpha)}}\Big\vert \leq 1,
$$
$$
\Big\vert \frac{p_C^a(\alpha) - \sum_\beta p_C^b(\beta) p^{a\vert
b}(\alpha\vert \beta)}{2 \sqrt{\prod_\alpha p_C^b(\beta) p^{a\vert b}(\alpha\vert \beta)}}\Big\vert \leq 1,
$$ 
\medskip

A context $C$ such that R3) holds is called trigonometric, because in this case we have the conventional formula of 
trigonometric interference: 
\begin{equation}
\label{TNCZ} p_C^b(\beta) = \sum_\alpha p_C^a(\alpha) p^{b\vert
a}(\beta\vert \alpha) + 2 \cos\theta(\beta\vert  \alpha, C)
\sqrt{\prod_\alpha p_C^a(\alpha) p^{b\vert a}(\beta\vert \alpha)},
\end{equation}
where $
\lambda(\beta\vert a ,C)=\cos \theta (\beta\vert a,C).
$
Parameters $\theta(\beta\vert \alpha,C)$ are said to be $b \vert
a$-{\it relative phases} with respect to the context $C.$ We defined these phases 
purely on the basis of probabilities. We have not started with any linear space; in contrast we shall
define geometry from probability.\footnote{We remark that conditions R1) and R3) are also nessesary.}

We denote the collection of all trigonometric contexts by the symbol ${\cal C}^{\rm{tr}}.$

By using the elementary formula:
$$
D=A+B+2\sqrt{AB}\cos \theta=\vert \sqrt{A}+e^{i \theta}\sqrt{B}|^2,
$$
for real numbers $A, B > 0, \theta\in [0,2 \pi],$ we can represent
the probability $p_C^b(\beta)$ as the square of the complex
amplitude (Born's rule):
\begin{equation}
\label{Born} p_C^b(\beta)=\vert \psi_C(\beta) \vert^2 \;.
\end{equation}
Here
\begin{equation}
\label{EX1} \psi(\beta) \equiv \psi_C(\beta)=
\sqrt{p_C^a(\alpha_1)p^{b\vert a}(\beta\vert \alpha_1)} + e^{i
\theta_C(\beta)} \sqrt{p_C^a(\alpha_2)p^{b\vert a}(\beta\vert
\alpha_2)}, \; \beta \in X_b,
\end{equation}
where $\theta_C(\beta)\equiv \theta(\beta\vert  \alpha, C).$

\medskip

The formula (\ref{EX1}) gives the quantum-like representation
algorithm -- QLRA. For any trigonometric context $C$ by starting
with the probabilistic data -- $ p_C^b(\beta), p_C^a(\alpha),
p^{b\vert a}(\beta\vert \alpha)$ -- QLRA produces the complex
amplitude $ \psi_C.$ This algorithm can be used in any domain of
science to create the QL-representation of probabilistic data (for a
special class of contexts).

We point out that QLRA contains the reference observables as
parameters. Hence the complex amplitude give by (\ref{EX1}) depends
on $a,b: \psi_C \equiv \psi_C^{b\vert a}.$

We denote the space of functions: $\varphi: X_b\to {\bf C}$ by the
symbol $\Phi =\Phi(X_b, {\bf C}).$ Since $X= \{\beta_1, \beta_2 \},$ the
$\Phi$ is the two dimensional complex linear space. By using QLRA
 we construct the map 
\begin{equation}
\label{TY}
J^{b \vert a}:{\cal C}^{\rm{tr}}
\to \Phi(X, {\bf C})
\end{equation}
 which maps contexts (complexes of, e.g.,
physical conditions) into complex amplitudes. The representation
({\ref{Born}}) of probability is nothing other than the famous {\bf
Born rule.} The complex amplitude $\psi_C(x)$ can be called a
{\bf wave function} of the complex of physical conditions (context)
$C$  or a  (pure) {\it state.}  We set $e_\beta^b(\cdot)=\delta(\beta-
\cdot)$ -- Dirac delta-functions concentrated in points $\beta=
\beta_1, \beta_2.$ The Born's rule for complex amplitudes (\ref{Born}) can be
rewritten in the following form: $\label{BH}
p_C^b(\beta)=\vert \langle \psi_C, e_\beta^b \rangle \vert^2,$ where the scalar product
in the space $\Phi(X_b, C)$ is defined by the standard formula:
$\langle \phi, \psi \rangle = \sum_{\beta\in X_b} \phi(\beta)\bar \psi(\beta).$ The system
of functions $\{e_\beta^b\}_{\beta\in X_b}$ is an orthonormal basis in the
Hilbert space $H_{ab}=(\Phi, \langle \cdot, \cdot \rangle).$ 

Let $X_b \subset {\bf R}.$ By using
the Hilbert space representation  of the Born's rule  we
obtain  the Hilbert space representation of the expectation of the
observable $b$: $E(b \vert C)= \sum_{\beta\in
X_b} \beta\vert\psi_C(\beta)\vert^2= \sum_{\beta\in X_b} \beta \langle \psi_C, e_\beta^b\rangle
\overline{\langle\psi_C, e_\beta^b\rangle}= \langle \hat b \psi_C, \psi_C\rangle,$ where
the  (self-adjoint) operator $\hat b: H_{ab} \to H_{ab}$ is determined by its
eigenvectors: $\hat b e_\beta^b=\beta e^b_\beta, \beta\in X_b.$ This is the
multiplication operator in the space of complex functions
$\Phi(X_b,{\bf C}):$ $ \hat{b} \psi(\beta) = \beta \psi(\beta).$ It is
natural to represent the $b$-observable  (in the Hilbert space
model)  by the operator $\hat b.$ 

We would like to have Born's rule
not only for the $b$-variable, but also for the $a$-variable:
$p_C^a(\alpha)=\vert \langle \varphi, e_\alpha^a \rangle\vert^2 \;, \alpha \in  X_a.$

How can we define the basis $\{e_\alpha^a\}$ corresponding to the
$a$-observable? Such a basis can be found starting with interference
of probabilities. We set $u_j^a=\sqrt{p_C^a(\alpha_j)},
p_{ij}=p(\beta_j \vert \alpha_i), u_{ij}=\sqrt{p_{ij}}, \theta_j=\theta_C(\beta_j).$ We
have:
\begin{equation}
\label{0} \varphi=u_1^a e_{\alpha_1}^a + u_2^a e_{\alpha_2}^a,
\end{equation}
where
\begin{equation}
\label{Bas} e_{\alpha_1}^a= (u_{11}, \; \; u_{12}) ,\; \; e_{\alpha_2}^a= (e^{i
\theta_1} u_{21}, \; \; e^{i \theta_2} u_{22})
\end{equation}
The condition R1) implies that the  system $\{e_{\alpha_i}^a\}$  is an orthonormal basis iff 
the probabilistic phases satisfy the
constraint:
$$
\theta_2 - \theta_1= \pi \; \rm{mod} \; 2 \pi,
$$  but, as we have seen \cite{KHC}, we can always choose such  phases (under the condition R1).

In this case the $a$-observable is represented by the operator
$\hat{a}$ which is diagonal with eigenvalues $\alpha_1,\alpha_2$ in the basis
$\{e_{\alpha_i}^a\}.$ The  conditional average of the observable
 $a$ coincides with the quantum Hilbert space average:
$
E(a \vert C)=\sum_{\alpha \in X_a} \alpha p_C^a(\alpha) = \langle \hat{a} \psi_C, \psi_C \rangle.
$

If condition R3) is violated, then we obtain nonconventional QL-representations of probabilistic
data, for example, in  the hyperbolic analogue of the complex Hilbert space \cite{KHC}.

It is important to remark that {\it map (\ref{TY}) is not one-to one!!!}
 Different contexts can be mapped into the same 
complex probability amplitude, we can also say that the same wave function may represent a few different contexts, 
cf. section 10.2.

Finally, we mention one recent result of Karl Svozil \cite{SV1}
which seems to be coupled to 
the  number of reference observables -- two -- producing QL-representation in our approach.
 
\section{Flyes in a packet}

We consider a metal box. At different points inside this box there is food which is attractive for flyes. 
Its distribution is not uniformly weighted, 
in some points there is more food than in others, there are domains without food. An external observer
(who is staying outside this box) has no idea about the real distribution of food in the box, but a ``Einstein
demon'' living inside this box knows well this distribution.  
We put a population of flyes, say $\Omega,$ 
inside this box. After while
they will be distributed in space inside the box by coupling to sites with food. 
Our  Einstein's demon   can find the probability distribution
${\bf P}(x,y,z)$  to observe a fly at the point with coordinates $(x,y,z).$ 
It is assumed to be stationary (at least for a while). In principle, some flyes can move between attractive 
points, but statistically the number of flyes at each site with food is stable.

As in the example ``firefly in the box'',
 one can divide this box in two ways: a) by the vertical wall -- $a,$ see Figure 2; 
b) by horizontal wall -- $b,$ see Figure 3. Here $a(\omega)=\alpha_1$ if  Einstein's demon  finds 
a fly $\omega$ in the left-hand part and $a(\omega)=\alpha_2$ if he finds 
a fly $\omega$ in the right-hand part (e.g. $\alpha=\pm 1).$
 We define $b$ in a similar way: $b(\omega)=\beta_1$ if  Einstein's demon  finds 
a fly $\omega$ in the top part and $b(\omega)=\beta_2$ if he finds 
a fly $\omega$ in the bottom part (e.g. $\beta = \pm 1).$ 
 Einstein's demon  can consider populations of flyes:
$$
\Omega_\alpha=\{\omega \in \Omega:
a(\omega)=\alpha\}, \; \; \Omega_\beta=\{\omega \in \Omega:
a(\omega)=\beta \}.
$$
By assuming that ${\bf P} (\Omega_\alpha), {\bf P}(\Omega_\beta) >0,$  he can define transition probabilities:
$$
p^{b \vert a}(\beta \vert \alpha) = {\bf P } (\Omega_\beta \vert \Omega_\alpha) \equiv
\frac{{\bf P}(\Omega_\beta \cap \Omega_\alpha)}{{\bf P}(A_\alpha)}
$$
and in the same way probabilities $p^{a \vert b}(\alpha\vert \beta).$

Let $C$ be some domain inside the box. We shall consider it as a geometric-context. 
 Einstein's demon  can be find (by using the Bayes' formula) conditional probability distribution:
\begin{equation}
\label{CPD}
{\bf P}_C(U)= \frac{{\bf P}(\Omega_U \cap \Omega_C )}{{\bf P}(\Omega_C)},
\end{equation}
for any subset $U$ of box.   
Here $\Omega_C = \{\omega \in \Omega: (x_\omega, y_\omega, z_\omega)  \in C\}$ is the populattion of flyes
which are concentrated inside the configuration $C.$ The population $\Omega_U$ 
is defined in the same way.

This probability distribution ${\bf P}_C$ provides the probabilistic representation of the domain $C.$
 Einstein's demon  encoded geometry by probability. Of course, probability provides only rough 
images of geometric structures, since 
the map: $$C \to {\bf P}_C$$ is not one-to-one. Denote now by ${\cal F}$ some $\sigma$-algebra 
of subsets of the box such that the probability
${\bf P}$ -- flyes' distribution -- can be defined on it. Denote also the set of all probability measures 
on the ${\cal F}$ by the symbol ${\cal P}.$  Then we have the map:
\begin{equation}
\label{CPD2}
J: {\cal F} \to {\cal P}.
\end{equation}
This is the classical probabilistic representation of geometry (of distribution of food). It is avalaible
for any internal observer ( Einstein's demon ) who lives inside this box. 
In this mapping a lot of geometric information is neglected. However, the whole probabilistic information 
is taken into account. This is the end of the classical story!

\medskip

{\bf Remark 6. 1.} (Food and flyes version of fields and particles) This representation has one interesting feature. Geometry of food distribution is represented by ensembles of flyes.
We can make the following analogy: electromagnetic field can be represented by photons.
One can compare the food distribution with a kind of a ``food-field'' and flyes with particles representing this field.
If we put another type of insects into the box, they may be not interested in this sort of food. They would not 
reproduce the distribution ${\bf P}(x,y,z).$ Thus we may speak about various food-fields which are represented
by corresponding types of insects-particles. In some sense this picture reminds Bohmian 
mechanics \cite{BM}.

\medskip

Now we modify the previous framework. We have the same box with the same ditribution of fly-attractive food.
But flyes are put not directly in the box, but in a plastic packet, say $C.$ The geometric 
configuration is unknown for us -- external observers. 
Moreover, we are not able to find its configuration directly (even by making a hole in the box), 
because packet's surface is covered by 
a ``B2-bomber type'' material. Thus we look inside the box, but we see nothing.\footnote{The ``Einstein
demon'' also gets a problem, but he can still investigate packet's geometry just by moving over its surface.
Of course, if the packet is disconnected, so it has a few components, a few ``Einstein
demons'' should be employed.} Nevertheless, we (external observers) would like 
to get at least partial information about this packet configuration by using flyes distribution. 
The problem is complified by the assumption that any attempt to open the metal box will induce
 destruction of the packet which in its turn induces  
redistribution of flyes in the space. Such a hard problem...

We do the following. As in the firefly-example we introduce fuzzy coordinates $a$ and $b.$ We measure 
them in the following way.
We assume that we can put very quickly either vertical or horyzontal  wall into the box. Such a moving wall
divides (pactically instanteneously, at least in  comparation with fly's velocity) 
the box into two sub-boxes, but at the same time it destroys (of course) the plastic packet. It is assumed that after 
this act we can open each sub-box and find numbers of flyes in each part of the box. 

At the moment, cf. section 8, we consider {\it nondisturbing measurements:} walls do not change food distributions 
in corresponding parts of the box (those  walls are negligibly thin and destruction of the packet does not change 
the distribution of food). However, opening of any box induces a strong disturbing effect, flyes are essentially 
redistributed.

Thus first we do the $a$-measuring by using the vertical wall. It divides the box
into two parts, say  $C_{\alpha_1}$ and $C_{\alpha_2}.$ In this way we get probabilities
$p_C^a(\alpha)$ that a fly was located  in the $\alpha$-side of the box. Since the vertical wall moves 
quicky relatively to fly's velocity, the number of flyes which were able to change the left-hand part 
of the box to the right-hand part or vice versa is statistically negligible. 
In principle, we might try to use the classical formula:
$$
p_C^a(\alpha)={\bf P }_C (\Omega_\alpha) \equiv  \frac{{\bf P}(\Omega_\alpha \cap \Omega_C)}{{\bf P}(\Omega_C)},
$$ 
However, it is tottaly unuseful for us, because we do not know the configuration $C$ and hence ${\bf P }_C.$

We point out that if we do not open sub-boxes $C_\alpha$ and 
if after while the corresponding ``Einstein demons'' measure
the $b$-coordinate of flyes in each part $C_\alpha$ 
of the box they will obtain the original transition probabilities 
$p^{b \vert a}(\beta \vert \alpha),$ since flyes will again redistribute in the domain $C_\alpha$ 
according to the food-field.\footnote{Two ``Einstein demons'' should be involved -- one for each sub-box.}
However, the original distribution of flyes in the 
domain $C \cap C_\alpha$ has been lost for ever even for  the ``Einstein demons.'' We (external observers) are not able 
to find transition probabilities in this way, since opening of a box produces redistribution of flyes 
in it.

We also remark that trivially $a(\omega)=\alpha$ on the $\alpha$-part of the box.

\medskip

{\bf Remark 6.2.} (Reaction of ``food-field'' to space reconfiguration)  At the moment we proceed under the assumption that 
the ``food-field'' is not sensitive to the disturbing effect of the moving wall (separating the box into two
sub-boxes). Moreover,   
the ``food-field'' is {\it not sensitive to changes of the geometry of space} (``boundary conditions''). 
In principle, we can imagine the following situation, see section 7.
The appearance of a separating wall does not induce a disturbing effect which 
could move food in space. However, the wall by itself can have some physical properties influencing the food 
distribution. For example, food is placed in charged capsulas and walls of the box 
(including walls used in  separation
experiments) also carry electric charges. Thus even ``mechanically peaceful appearance'' of a separating wall will induce
(after a while) redistribudtion of food in the sub-box.       

\medskip

{\bf Remark 6.3.} (Fair sampling) At the moment we proceed under the assumption of ``fair sampling,'' cf. 
\cite{Santos}, \cite{AD}, \cite{AD1}.  
Moving walls do not kill statistically non-negligible populations of flyes.

\medskip

To construct the QL-representation  of the context $C$ by a complex  probability amplitude, we need
also probabilities:
$$
p_C^b(\beta)={\bf P }_C (\Omega_\beta) \equiv  \frac{{\bf P}(\Omega_\beta \cap \Omega_C)}{{\bf P}(\Omega_C)},
$$ 
However, since we do not know the configuration $C,$ we are not able to apply Bayes' formula directly.
We should repeat previous considerations, but by using now the horizontal wall which separates 
quickly the box into top and bottom parts, $C_{\beta_1}$ and $C_{\beta_2}.$
Then by opening these sub-boxes and counting flyes in each  of them 
we find the probabilities $p_C^b(\beta).$ 

Of course, we should have two boxes with the same configuration $C,$ because each falling wall destroys 
this configuration. Thus we should be able to make such a preparation a few times. Moreover, if one wants
to exclude effects of interaction between flyes (as one does in QM), there should be created an ensemble of boxes,
each box containing just one fly. It is assumed that flyes would reproduce the food distribution.

In particular, for $C=C_\alpha,$ i.e., the configuration $C$ which coinsides with the $\alpha$-part of the box
we get: $\Omega_{C_\alpha}=\Omega_\alpha$ and 
$$
p_{C_\alpha}^b(\beta)= p^{b \vert a}(\beta \vert \alpha).
$$ 
However, we do not know from the very beginning that a hidden geometric configuration is the half-box $C_\alpha.$
Therefore this is not an experimental way to find transition probabilities. 

To find transition probabilities, we assume that each half-box $C_\alpha$ can be devided by the horyzontal wall
(as in the original $b$-measurement in the whole box) in two parts, say $C_{\beta \vert \alpha}, \beta=\beta_1, 
\beta_2.$ By counting flyes in each of these boxes we find the transition probabilities. At the moment we proceed
under the same assumptions as before: by puting the horyzontal walls in the box $C_\alpha$ 
we do not change the distribution of food in it.

Now everything is prepared for application of QLRA.
A nessesary condition is given by R2), since 
in QM matrices of transition probabilities are symmetrically conditioned.
Thus from the very beginning one should assume that the distribution of attracting sites in the box induces 
this condition. This happens iff ${\bf P}(\Omega_\alpha)={\bf P}(\Omega_\beta)=1/2.$

The next condition is that variables are statistically conjugate, i.e.,
${\bf P}(\Omega_\alpha\cap \Omega_\beta)\not=0$ for all $\alpha$ and $\beta.$ 

Finally, the context $C$ should be ``large enough'' with respect to both variables: 
${\bf P}(\Omega_C \cap \Omega_\beta),  {\bf P}(\Omega_C \cap \Omega_\alpha)>0.$
Statistically small configurations could not be represented in the QL-way (they
are simply neglected in the incomplete QL-representation of information).

We also know that, becides a complex probability amplitude, some contexts can be represented by hyperbolic
amplitudes, thus to guarantee real QM-like representation we should have $\vert \lambda\vert \leq 1$ for the 
coefficient of interference.
 
Thus  we represent all ``trigonometric configurations'' $C$ by complex vectors and the observables
$a$ and $b$ by self-adjoint operators. The map:
$$
J^{b\vert a}: {\cal C}^{\rm{tr}} \to H
$$
is a QL-analogue of the classical map $J$ given  by (\ref{CPD2}). Of course, 
the map (\ref{CPD2})  is ``better'' than the QL-map.
However, we are not able to use it in the situation with invisible configuration $C.$

As was remarked, for some contexts, hyperbolic ones, $\vert \lambda\vert > 1.$ They are
 mapped into hyperbolic amplitudes: 
$$
J^{b\vert a}: {\cal C}^{\rm{hyp}} \to H_{\rm{hyp}},
$$ 
where $H_{\rm{hyp}}$ is the hyperbolic analogue of Hilbert space \cite{KHC}.
Appearance of such amplitudes is not surprising from the viewpoint of general contextual probability theory.
Why should the coefficient $\lambda$ be always bounded by one?
It is surprising that we do not have them in conventional QM. In some way it
 happens that all physical quantum contexts 
are   trigonometric (or that physical hyperbolic contexts have not yet beeen observed?).

\medskip

{\bf Remark 6.4.} (Complementarity or supplementarity?) We point out that, although the ``reference observables''
$a$ and $b$  corresponding to two different separations of the box are represented 
by noncommutative operators, they can be considered as simutaneously
existing: each fly has the definite position in the box and hence its location in each part of the 
box is well defined. This is the typical situation for the Kolmogorov approach: the values of both random variables
$a(\omega)$ and $b(\omega)$ are well defined for each $\omega \in \Omega.$ Thus ``properties'' $a$ and $b$ 
of a fly are not mutually exclusive, in spite of noncommutativity: $[\hat{a}, \hat{b}]\not= 0.$
Since Nils Bohr reserved the term complementarity for mutually exclusive properties, it might be better 
to call $a$ and $b$ {\it supplementary observables}, see \cite{SP}. It is clear that a result of  measurement 
of $b$ produces supplementary information with respect to the result of preceding measurement of $a$ and vice versa.

\section{How far can one proceed with the quantum-like representation of the Kolmogorov model?}

In spite of the presence of the underlyning Kolmogorov space, we constructed the QL-representation of probabilistic data
for macroscopic configurations (essentially incomplete representation) which has all distinguishing features
of the conventional quantum representation of probabilistic data for a pair of incompatible
observables: intereference formula for probabilities, Born's rule, representation 
of these observables by self-adjoint operators. As was mentioned, the map  $J^{b\vert a}$ is not injective.
We no ask: Is it surjective? Can one get any quantum state $\psi$ and any pair of quantum observables
$\hat{a}$ and $\hat{b}$ in such a way? The answer is no. This is a consequence of Bell's type inequality for
transition probabilities, see \cite{CBI} and  appendix.

To apply conditional Bell's inequality to our macroscoipic situation, it is better 
to consider a ball bounded by the metal sphere, instead of the box. We now can divide this ball into parts
with the aid of central planes. To simplify considerations, we can consider a boundle of planes which are enumerated
by the angle $\phi.$ Then we shall obtain a familty of observables $a_{\phi},$ say taking values $\pm .$
Parts of the ball obtained by the $\phi$-separation are 
$C_{\phi,+}=\{ \theta: \phi \leq \theta < \phi+\pi\}$
and $C_{\phi,-}=\{ \theta: \phi+\pi \leq \theta < \phi\},$ respectively.
 
For each pair of them we find transition 
probabilities $p^{\phi_1 \vert \phi_2} (\epsilon_1 \vert \epsilon_2).$ 
For each context $C$ (a plastic packet with flyes inside it; this paket is placed inside the metal ball;
any attempt to open the ball would destroy this paket) and any $\phi$-section, we find probabilities  
$p_C^{\phi}(\epsilon), \epsilon= \pm .$
If we choose  a context $C$ such that    
$p_C^{\phi}(+1)=p_C^{\phi}(-1)=1/2$ for all $\phi,$ then we can apply arguments of appendix and we see that
some types of transition probabilities could not be obtained from a single Kolmogorov model. 

One Kolmogorov space is too small to generate all quantum (or better to say quantum-like)  states and observables.

\section{Disturbing measurements}

However, we can easily modify our example to destroy the (hidden) Kolmogorov structure of the model.
Suppose now that everything is as it was before with only one difference: destruction of the packet 
by a wall (encoded by some $\phi$-plane) induces not only the possibility for flyes to move outside the packet,
but also induces a redistribution of food sites, cf. Remark 2. 
The latter is determined by the wall. Thus after e.g. the $\phi$-plane separation of the ball
the distribution of sites with food in its parts $C_{\phi,+}$ and $C_{\phi,-}$ is not such as it was
before this separation.   Therefore, for any successive $\phi^\prime$-separation of the sectors $C_{\phi,\epsilon}$  
(which were produced by the previous $\phi$-separation), the transition probabilities 
$p^{\phi^\prime \vert \phi} (\epsilon^\prime \vert \epsilon)$ obtained
by an external observer do not coincide with the transition probabilities which would be obtained by 
 Einstein's demon  on the basis of the original ensemble. 
Hence Bell's type inequality for transition probabilities, see appendix, cannot be applied. 

In fact, by using random generators we can simulate  probabilities for any complex 
probability amplitude and any pair 
of self-adjoint operators in the two dimensional Hilbert space. 

For example, suppose that we would like to simulate the transition probabilities for successive 
measurements of spin projections as well as the uniform probability 
distribution for the $a_\phi$
measurements for the original context $C$ (state $\psi_C).$ To provide the latter condition, we start with
the uniform distribution of food. It would induce probabilities $p_C^{\phi}(+1) =
p_C^{\phi}(-1)=1/2.$

Now to simplify considerations, we consider not three dimensional configurations, but just two dimensional, 
in particular, we consider a circle, instead of a ball, and sections by central lines, instead of planes. 

We assume that disturbance induced by the $a_{\phi_0}$-measurement, $0 \leq \phi_0 < \pi,$ 
induces redustribution of food in the sectors $C_{\phi_0,+}$ and $C_{\phi_0,-}$ 
and, finally, generates e.g. in the  sector $C_{\phi_0,+}$ 
the density of flyes:
\begin{equation}
\label{DEN}
\rho_{\phi_0}^+ (r, \theta ) = \sin (\theta - \phi_0).
\end{equation}   
(We assume that the circle has unit radius).
Then we separate the sector $C_{\phi_0,+}$ by the $\phi$-plane, say $\phi > \phi_0.$ Then the probability
$$
p^{\phi \vert \phi_0}(+ \vert +)= \int_{0}^1 r d r\int_{\phi}^{\phi_0 + \pi} \sin (\theta - \phi_0)d \theta=
\cos^2 \frac{\phi - \phi_0}{2},
$$
$$
p^{\phi \vert \phi_0}(- \vert +)= \int_{0}^1 r d r \int_{\phi_0}^{\phi} \sin (\theta - \phi_0) d \theta=
\sin^2 \frac{\phi - \phi_0}{2}.
$$
For the sector $C_{\phi_0,-},$ we choose the probability distribution
\begin{equation}
\label{DEN1}
\rho_{\phi_0}^- (r, \theta ) = - \sin (\theta - \phi_0).
\end{equation}
Here transition probabilities are given by
$$
p^{\phi \vert \phi_0}(+ \vert -)= -\int_{0}^1 r d r\int_{\phi_0+ \pi}^{\phi + \pi} \sin (\theta - \phi_0)d \theta=
\sin^2 \frac{\phi - \phi_0}{2},
$$
$$
p^{\phi \vert \phi_0}(- \vert -)= - \int_{0}^1 r d r \int_{\phi+ \pi}^{\phi_0 + 2\pi} 
\sin (\theta - \phi_0) d \theta=
\cos^2 \frac{\phi - \phi_0}{2}.
$$

{\bf Remark 8.1.} (Complementarity or supplemntarity?) Since we consider disturbing measurements, we 
(external observers) are not able to measure
two observables, $a_{\phi_1}$ and $a_{\phi_2},$ simultaneously. Thus these are {\it incompatible observables.}
However, such measurement incompatibility does not exclude that an element of reality can be assigned to
each fly  -- the pair $a_{\phi_1}(\omega), a_{\phi_2}(\omega).$ We recall that we consider such separations that they do not induce 
redistribution of flyes between sectors: the $\phi$-plane moves so quickly that flyes are not able to change sectors
(or at least only statistically negligible number of flyes could make such changes). Moreover, only negligible 
number of flyes can be killed by a moving-separating plane. Thus the values of $a_{\phi_1}(\omega)$ 
and    $a_{\phi_2}(\omega)$ which would be obtained by an external observer coincide with the values which
have been  known by 
 Einstein's demon  before measurements. Therefore complementarity (in the sense of mutual exclusivity) is only external 
observer's complementarity.  Einstein's demon  still has supplementarity, in the sense of additional 
information (of course, fuzzy) about fly's location. 

\medskip

{\bf Remark 9.1.} (Counterfactual arguments) We point out that already in the previous remark we have applied
counterfactual arguments -- by using ``would be obtained by an external observer.'' In fact, one cannot escape
them, because an external observer is not able to assign both values  $a_{\phi_1}(\omega), a_{\phi_2}(\omega)$ to the same fly 
$\omega.$

\section{Can the classical probabilistic structure be violated without disturbance effects?}

In section 7 we pointed out that by Bell's inequality for transition probabilities it is impossible  
to find a single underlying classical probabilistic space which would
reproduce all possible wave functions and pairs of self-adjoint
noncommutative operators in the contextual probabilistic framework.
One can not find such a Kolmogorov probability space that by choosing different pairs of reference observables
$a,b$ and corresponding families of trigonometric contexts ${\cal C}^{\rm{tr}}(a,b)$ (represented by sets from the $\sigma$-algebra 
of the Kolmogorov space) he would (by alpplying QLRA)  cover the whole unit sphere of Hilbert state space as well as 
obtain all pairs of 
noncommutative self-adjoint operators. 
 In section 8 we showed that by considering
disturbing measurements we can reproduce all quantum structures. Can one approach the same result 
without disturbance? In principle, yes!

\subsection{Unfair sampling}

One of possibilities is to proceed under unfair sampling assumption, see
Remark 6.3, cf. \cite{Santos}, \cite{AD}, \cite{AD1}. We can assume that moving planes separating the metal ball do not produce 
food redistribution, thus the ``food-field'' is not changed. However, these planes kill 
subensembles of flyes\footnote{We remark that we really consider unfair samplig and not ``the detectors 
efficiency.'' We operate with macroscopic systems -- flyes which are are detected with probability one.}, 
depending on $\phi.$ Then we can easily violate the Bell's inequality for conditional
probabilities.

\subsection{Ensemble fluctuations}

Another important point is that in section 7 we proceeded by using counterfactual arguments, cf. remark 9.1.
To be on really realistic ground, we should consider at least three different balls and perform on them 
conditional measurents for pairs of observables $a_{\phi_i}, a_{\phi_j}.$ In principle, we cannot guarantee
that we would be able to reproduce statistically identical distributions of food in balls and idential 
hidden configurations. As was emphasized in section 3, the map, see (\ref{TY}), from the collection of trigonometric
contexts into complex probability amplitudes is not injection, various contexts can be mapped in the same
complex probability amplitude. Even if we are sure that we have the same QL-state given by the same 
complex probability amplitude, $\psi,$ we could never be sure that contexts in different 
balls are the same. It may be that $\psi=\psi_{C_1}=\psi_{C_2}=...=\psi_{C_N}$ and moreover
it may be that $N\to \infty.$
Therefore we should work in multi-Kolmogorovian framework and the Bell's inequality for conditional
probabilities can also be violated without any disturbance.

This argument (but for composite systems) was presented at the first time by De Baere \cite{DEB}, 
then by the author \cite{KH3} 
and recently by Hess and Philipp \cite{HPL2}. Moreover, they pointed out in \cite{HPL2}
to the old paper of Soviet 
mathematician Vorobjev \cite{VR} who studied the problem of the possibility to realize a number of observables
on a single Kolmogorv space. This problem is equivalent to the problem of violation of Bell's inequality 
for transition probabilities.  

\subsection{Communication}

In principle, we may also produce redistribution of flyes without redistribution of the ``food-field'' 
if we assume that flyes can communicate. For example, each separation measurement starts communication 
between flyes. As the result, they can come to the agreement to concentrate in each sector $C_{\phi_0, \pm}$
in such a way that e.g. $\sin$-type dsitribution of section 8 would be produced. If they communicate by using
signals which we, external observers, are not able to detect, then this communication would be hidden from us. 
  
\section{ERP-Bohm type experiments with flyes}

We have considered in very detail measurements (in fact, position-type measurements) for 
ensembles of single flyes. In principle, we could consider the real EPR-Bohm type experiment
for pairs of ``entangled flyes'' which we put into different metal balls. One of technological problems is to produce such pairs of flyes. 
However, this is not the main point. The main point is that in the macroscopic framework such 
experiments would not give so much more than experiments with single flyes. In contrast to photons or electrons,
we have no doubts that flyes have objective properties, in particular, the position. Therefore the only 
consequence of the EPR-Bohm type experiment with flyes would be that  disturbing effects should be 
excluded.\footnote{We remind that we consider not only mechanical disturbance by moving planes, but also the field type disturbance.
To exclude the latter type of disturbance, one should be sure that the effect of the ``food-field'' (e.g. smell) from one ball
would be not able to propagate to another ball. If balls have small windows (or produced not of metal, but of some less
isolating material), then smell can propagate from one ball 
to another. We recall that insects can find smell-traces on huge distances. Thus to exclude completely 
disturbing effects, we should either isolate balls completely or to make measurements on balls with 
a time-window such that a signal from one ball would not be able to approach another during this 
time window, cf. \cite{WH}}

Thus as well as in the case of a single system we have tree choices: 
a) unfair sampling; b)  ensemble fluctuations; c) nonrelativistic communications between flyes. 

The last condition cannot be completely rejected even for human beings, but the EPR-type experiment
could not be used to provide the crucial argument in its favor.

{\bf Conclusion.} {\it We shown that macroscopic configurations can be naturally represented in the
QL-way -- by complex probability amplitudes -- with the aid of pairs of ``supplementary observables''
which in turn are represented by noncommutative self-adjoint operators. Classical probabilistic structure
can be violated. In particular, Bell's type inequality can be violated. Such violations have nothing to do
with ``death of reality.'' They could be induced either by disturbing effects of measurements, or 
unfair sampling, or ensemble fluctuations, or nonlocal communication between macroscopic systems.
The latter assumption is not so much reasonable for macroscopic biological systems (however, it could not be completely 
excluded).}
   
\section{Appendix: Bell's inequality for transition probabilities}

{\bf Theorem.} {\it Let $a, b, c=\pm 1$ be
dichotomous uniformly distributed random variables on a single Kolmogorov space. Then 
the following inequality holds true:}
\begin{equation}
\label{BB1}
{\bf P}(a=+1 \vert b=+1) + {\bf P}(c=+1 \vert  b=-1) \geq {\bf P}(a=+1 \vert  c=+1)
\end{equation}

{\bf Proof.} We have 
$$
{\bf P}(b=+1)={\bf P}(b=-1) = {\bf P}(a=+1)={\bf P}(a=-1)={\bf P}(c=+1)={\bf P}(c=-1)=1/2.
$$
Thus
$$
{\bf P}(a=+1 \vert b=+1) + {\bf P}(c=+1 \vert b=-1)=2{\bf P}(a=+1, b=+1) + 2{\bf P}(c=+1, b=-1)
$$
and
$$
{\bf P}(a=+1 \vert c=+1)=2{\bf P}(a=+1, c=+1).
$$
Hence by the well know Wiegner  inequality \cite{WW} we get (\ref{BB1}).

\medskip

We underline again that the main distinguishing feature of (\ref{BB1}) is the presence
of only transition probabilities. Transition probabilities can always be calculated by using quantum
formalism for noncomposite systems. In fact, we need not consider pairs of particles.

\medskip

I would like to thank L. Accardi, A. Aspect,   A. Grib, E. Haven, G. `t Hooft, A. Leggett, 
S. Gudder for discussions on the possibility to apply the quantum formalism to macroscopic systems.

This paper was written during author's visiting professor fellowship (supported by DFG) at 
university of Bonn. I would like to thank Sergio Albeverio for hospitality and many years of supporting 
my investigations on quantum foundations. The results of this paper were presented at the General 
Seminar on Stochastics, Bonn University. I would like to thank all its participants for debates and advices.

\end{document}